\documentclass{article}
\usepackage{ijcai19}
\usepackage{times}
\usepackage{soul}
\usepackage{url}
\usepackage[hidelinks]{hyperref}
\usepackage[utf8]{inputenc}
\usepackage[small]{caption}
\usepackage{graphicx}
\usepackage{amsmath}
\usepackage{booktabs}
\usepackage{algorithm}
\usepackage{algorithmic}
\usepackage{microtype}
\usepackage{graphicx}
\usepackage{subcaption}
\usepackage{booktabs} 
\usepackage{hyperref}
\usepackage{url}
\usepackage{color}
\usepackage{amstext} 
\usepackage{multirow}
\usepackage{mathrsfs}
\usepackage{amsmath}
\usepackage{nth}
\usepackage{amsfonts}
\usepackage{mathtools}
\usepackage{stmaryrd}
\usepackage{footnote}
\usepackage{siunitx} 
\usepackage{paralist}

\title{STAR-GCN: Stacked and Reconstructed Graph Convolutional Networks for \\Recommender Systems}

\author{}

\author{
Jiani Zhang$^1$ \and
Xingjian Shi$^2$ \and
Shenglin Zhao$^3$
\And
Irwin King$^1$
\affiliations
$^1$ The Chinese University of Hong Kong, Hong Kong, China\\
$^2$Hong Kong University of Science and Technology, Hong Kong, China\\
$^3$Youtu Lab, Tencent, Shenzhen, China\\
\emails
\{jnzhang, king\}@cse.cuhk.edu.hk,
xshiab@connect.ust.hk, zsl.zju@gmail.com
}

\begin{document}

\maketitle
\begin{abstract}
We propose a new \emph{\underline{STA}cked and \underline{R}econstructed \underline{G}raph \underline{C}onvolutional \underline{N}etworks}~(STAR-GCN) architecture to learn node representations for boosting the performance in recommender systems, especially in the cold start scenario. STAR-GCN employs a stack of GCN encoder-decoders combined with intermediate supervision to improve the final prediction performance. Unlike the graph convolutional matrix completion model with one-hot encoding node inputs, our STAR-GCN learns low-dimensional user and item latent factors as the input to restrain the model space complexity. 
Moreover, our STAR-GCN can produce node embeddings for new nodes by reconstructing masked input node embeddings, which essentially tackles the cold start problem. 
Furthermore, we discover a label leakage issue when training GCN-based models for link prediction tasks and propose a training strategy to avoid the issue.
Empirical results on multiple rating prediction benchmarks demonstrate our model achieves state-of-the-art performance in four out of five real-world datasets and significant improvements in predicting ratings in the cold start scenario. The code implementation is available in \url{https://github.com/jennyzhang0215/STAR-GCN}.
\end{abstract}

\section{Introduction}
Recommender systems, which analyze users' preference patterns to suggest potential targets, are indispensable in content providers, electronic retailers, web search engines, etc. The key mathematical problem underlying recommender systems is matrix completion~\cite{candes2009exact}. Assume there are $N$ users and $M$ items, the recommendation algorithm aims to fill in the missing entries in the $N \times M$ rating matrix given the existing entries.

The classical way to solve this problem is via \emph{Matrix Factorization} (MF)~\cite{koren2009matrix}, in which the rating scores are generated by functions over the latent factors or embeddings of users and items. Recent advancements in deep learning, especially \emph{Graph Convolutional Networks} (GCN)~\cite{defferrard2016convolutional,bronstein2017geometric,kipf2017semi,hamilton2017inductive}, have brought new ideas for tackling this essential artificial intelligence problem. GCN generalizes the definition of convolution from the regular grid to irregular grid, like graph structures. 
The GCN framework generates node representations by a localized parameter-sharing operator, known as {\emph graph aggregator}~\cite{hamilton2017inductive,zhang2018gaan}. A graph aggregator calculates a node's representation by transforming and aggregating the features of its local neighborhoods. By stacking multiple graph aggregators and nonlinear functions, we build a deep neural network that can extract features across far reaches of a graph. Because the local neighborhood set can be viewed as the receptive field of a convolution kernel, this kind of neighborhood aggregation methods is named as \emph{graph convolution}, which also have connections to spectral graph theory~\cite{kipf2017semi}.

Monti et al.~\shortcite{monti2017geometric} proposed the first GCN-based method for recommender systems. In their approach, GCN was used to aggregate information from two auxiliary user-user and item-item graphs. The latent factors of users and items were updated after each aggregation step, and a combined objective function of GCN and MF was used to train the model. After that, Berg et al.~\shortcite{berg2017graph} proposed the \emph{Graph Convolutional Matrix Completion} (GC-MC) model. GC-MC directly characterized the relationship between users and items as a bipartite interaction graph. Two multi-link graph convolution layers were used to aggregate user features and item features. The ratings were estimated by predicting the edge labels. Thanks to the power of GCN in learning high-quality user and item representations, GC-MC has achieved state-of-the-art performance in several public recommendation benchmarks.

While being powerful, the GC-MC model has two significant limitations. To distinguish each node, the model uses one-hot vectors as node input. This makes the input dimensionality proportional to the total number of nodes and thus is not scalable to large graphs. Moreover, the model is unable to predict the ratings for new users or items that are not seen in the training phase because we cannot represent unknown nodes as one-hot vectors.
 The task of predicting ratings for new users or items is also known as the \emph{cold start} problem.

In this paper, we propose a new architecture, \emph{\underline{STA}cked and \underline{R}econstructed \underline{G}raph \underline{C}onvolutional \underline{N}etworks} (STAR-GCN), to solve these problems. 
Unlike GC-MC, STAR-GCN directly learns low-dimensional user and item embeddings as the input to the network in an end-to-end fashion. To improve the learned embeddings and also generalize the model to predict embeddings of unseen nodes for the cold start problem, STAR-GCN masks a part of or the whole user and item embeddings and reconstructs these masked embeddings with a block of graph encoder-decoder in the training phase. This technique is inspired by the recent success of the `masked language model' in learning language embeddings~\cite{devlin2018bert}. Moreover, we build a stack of encoder-decoder blocks in conjunction with intermediate task-specific supervision to enhance the final performance. 
During implementation, we find that training the GCN-based models for rating prediction faces the label leakage issue, which results in the overfitting problem and significantly degrades the final performance. To avoid the leakage issue, we provide a \emph{sample-and-remove} training strategy and empirically demonstrate the effectiveness.

We conduct experiments over two tasks: transductive rating prediction and inductive rating prediction. The transductive rating prediction is generally used in traditional matrix completion tasks, i.e., all the testing users and items are observed in training data. The inductive rating prediction is a newly introduced task to evaluate different models' ability on the cold start problem. We ask new users to rate a few items or require new items to be rated by a few users. These data are only used in the inference step to elicit initial information about new users/items, which is similar to the ask-to-rate technique~\cite{nadimi2014cold} for cold start.
Experiments show that STAR-GCN achieves state-of-the-art performance in four out of five real-world datasets in the transductive setting. In the inductive setting, our STAR-GCN  consistently and significantly outperforms the baselines.

Our main contributions are: (1) we propose a new architecture for recommender systems to learn latent factors of users and items in both transductive and inductive settings; (2) we are the first to explicitly pinpoint a training label leakage issue when implementing GCN-based models in rating prediction tasks and propose a training strategy to avoid this issue, leading to substantial performance improvement; (3) our STAR-GCN models achieve state-of-the-art performance in four out of five real-world recommendation datasets in the transductive setting and significantly outperform other models in the inductive setting.

\section{Preliminary}
We denote vectors with bold lowercase letters, matrices with bold uppercase letters, and sets with calligraphy letters. We omit the bias variable of linear transformation for brevity.

\subsection{Rating Prediction Tasks}
\begin{figure}
    \centering
    \begin{subfigure}[b]{0.485\linewidth}
        \centering
        \includegraphics[width=0.95\textwidth]{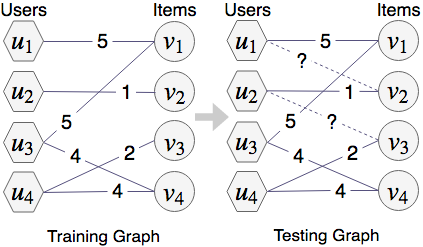}
        \caption{Transductive rating prediction}
        \label{fig:pre_trans}
    \end{subfigure}
    ~
    \begin{subfigure}[b]{0.485\linewidth}
        \centering
        \includegraphics[width=0.95\textwidth]{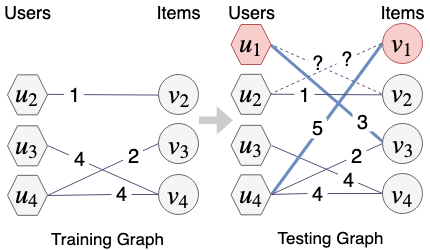}
        \caption{Inductive rating prediction}
        \label{fig:pre_ind}
    \end{subfigure}
    \vspace{-0.4cm}
    \caption{Two rating prediction tasks. (a) All users and items are observed in the training graph. (b) $u_1$ and $v_1$ are not seen in the training time. But we can access a few edges connected with the new nodes in the testing graph before we make predictions.}
    \label{fig:task}
    \vspace{-0.2cm}
\end{figure}

GCN-based models treat the recommendation environment as an undirected bipartite graph $G$ that contains two disjoint node sets, users $\mathcal{U}$ and items $\mathcal{V}$. Suppose there are $N$ users and $M$ items, an edge value $r_{i,j} \in \mathcal{R}$ represents an observed rating value from user $u_i$ to item $v_j$. The rating set $\mathcal{R}$ may contain several ordinal rating levels, i.e., $r\in\{1,...,R\}=\mathcal{R}$. Each rating level indicates a link type in the bipartite graph. All the training rating pairs form a training graph, and they are included in a testing graph. Examples are shown in Figure~\ref{fig:task}.
The goal of rating prediction is to predict the ratings a user would give to other items, given a small subset of observed rating pairs. We focus on two types of rating prediction tasks: transductive rating prediction and inductive rating prediction. Figure~\ref{fig:task} illustrates the difference between these two tasks.

\textbf{Transductive Rating Prediction}. Figure~\ref{fig:pre_trans} shows a transductive rating prediction example, where users and items appearing in the testing graph are observed in the training graph. Prior collaborative filtering methods, like \emph{Matrix Factorization} (MF)~\cite{koren2009matrix}, primarily concentrate on this task.

\textbf{Inductive Rating Prediction}. Figure~\ref{fig:pre_ind} illustrates an example of the inductive rating prediction task. $u_1$ and $v_1$ are two new nodes, which are not seen at the training time but appear in the testing rating pairs. Before making predictions, we access a few rated edges connected with these new nodes in the testing graph. Traditional collaborative filtering methods cannot solve this task without re-training the models. The standard way is to rely on content information to model the users' and items' preferences. The \emph{Collaborative Deep Learning} (CDL)~\cite{wang2015collaborative} model and \emph{DropoutNet}~\cite{volkovs2017dropoutnet} are two recent representative methods. The core idea of these two models is to use a deep neural network to learn effective features of the input node content. 

Recent progress in deep learning on graphs, mainly the GCN models, can address the above two tasks by learning transductive and inductive node representations. STAR-GCN inherits the ability of GCN for both transductive and inductive learning. Compared with CDL and DropoutNet, our STAR-GCN not only takes account of the node's content information but also utilizes the structural information to learn the embeddings of new nodes. Thus, STAR-GCN can solve the cold start problem when the content information is unavailable, which is infeasible for CDL and DropoutNet.

\begin{figure*}
    \centering
    \begin{subfigure}[b]{0.365\linewidth}
        \centering
        \includegraphics[width=0.65\textwidth]{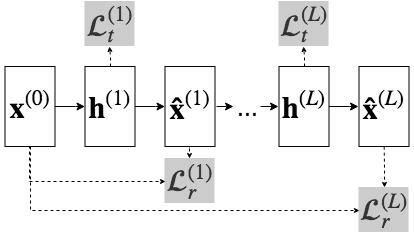}
        \caption{Our STAR-GCN architecture with $L$ blocks}
        \label{fig:block_recon}
    \end{subfigure}
    ~
    \begin{subfigure}[b]{0.395\linewidth}
        \centering
        \includegraphics[width=0.63\textwidth]{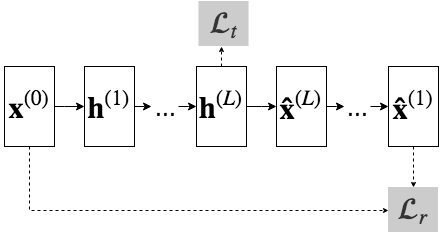}
        \caption{One block with $L$-layer encoder and decoder}
        \label{fig:layer_recon}
    \end{subfigure}
    ~
    \begin{subfigure}[b]{0.21\linewidth}
        \centering
        \includegraphics[width=0.67\textwidth]{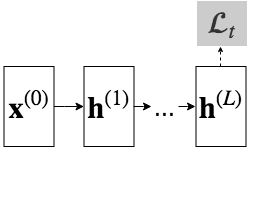}
        \caption{Without reconstruction decoder}
        \label{fig:block_norecon}
    \end{subfigure}
    \vspace{-0.5cm}
    \caption{Model Architectures. $\mathbf{x}^{(0)}$ is the initial node vector. $\mathbf{h}^{(\ell)}$ is the output of the $\ell$-th graph encoder and $\mathbf{\hat{x}}^{(\ell)}$ is the output node vector of the $\ell$-th decoder. $\mathcal{L}_t$ is a task specific loss and $\mathcal{L}_r$ is a reconstruction loss.  (b) and (c) are two special cases of (a).}
    \label{fig:model}
    \vspace{-0.2cm}
\end{figure*}

\subsection{Graph Convolutional Matrix Completion}
In this subsection, we briefly revisit GC-MC~\cite{berg2017graph}. Our STAR-GCN employs a similar graph aggregator to encode structural information.  
GC-MC uses a multi-link graph convolutional encoder to produce node representations. Each link type $r$ is assigned with a specific transformation. The messages from items to user $u_i$ are computed as
\vspace{-0.1cm}
\begin{equation}
    \begin{aligned}
    \mathbf{y}_{u_i}^r &= \sum_{v_j \in \mathcal{N}_r(u_i)} \frac{1}{c_{ij}}\mathbf{W}_{a}^r \mathbf{x}_{v_j},\\
    \mathbf{h}_{u_i} &= \alpha(\mathbf{W}_{h}\alpha(\sum_{r=1}^R\mathbf{y}_{u_i}^r)).
    \end{aligned}
    \label{eq:gcmc}
\vspace{-0.2cm}
\end{equation}
Here, $\mathbf{y}_{u_i}^r$ is the aggregated output of a link type $r$. $\mathbf{x}_{v_j} \in \mathbb{R}^{d_{in}}$ is the initial vector of $v_j$ and $d_{in}$ is the input dimension. $\mathbf{W}_{a}^r$ of size $d_a\times d_{in}$ is a link-specific weight matrix for rating level $r$, which transforms a vector of the dimension $d_{in}$ to an hidden size $d_a$. $c_{ij}$ is a normalized constant, computed as $\sqrt{|\mathcal{N}_r(u_i)||\mathcal{N}_r(v_j)|}$ with $\mathcal{N}_r(\cdot)$ denoting a set of neighbors connected by edge-value $r$. After computing the link specific messages, we sum the messages from total $R$ types of links and pass the output to a non-linear function $\alpha(\cdot)$. Finally, we employ a fully connected layer with parameter $\mathbf{W}_{h}$ of size $d_h\times d_a$ and another non-linear activation to produce the final node vector $\mathbf{h}_{u_i}$ for user $u_i$. Messages from users to items are processed analogously with a separate set of parameters.

In the next step, the GC-MC model takes the computed user vector $\mathbf{h}_{u_i}$ and item vector $\mathbf{h}_{v_j}$ as the input to predict the rating value $\hat{r}_{i,j}$. See Berg et al.~\shortcite{berg2017graph} for more details.
In the following section, we do not distinguish user $u$ and item $v$ and call them as a node.

\section{Our Models}

The architecture of STAR-GCN is a multi-block graph encoder-decoder shown in Figure~\ref{fig:block_recon}. The multi-block architecture allows reassessment of initial estimates and features across the whole graph. 
In particular, each block contains two components: a graph encoder and a decoder. The graph encoder generates node representations by encoding semantic graph structures as well as input content features, and the decoder aims to recover the input node embeddings. For each block, we impose a task-specific loss after graph encoders and a node reconstruction loss after decoders.

STAR-GCN supports two different types of combinations between two consecutive blocks, by stacking or by recurrence. The main difference is whether to share parameters among blocks or not. By stacking, we consecutively place multiple encoder-decoder blocks with separate sets of parameters. By recurrence, we unfold a single encoder-decoder block, so the same set of parameters are shared across all the blocks, which curtails the total memory usage. 
Besides, our STAR-GCN is a general framework, which can be simplified to some individual cases, as shown in Figure~\ref{fig:layer_recon} and~\ref{fig:block_norecon}. Empirical studies on two combinations and the simplified models are in Table~\ref{tb:ablation}.

\subsection{Input Node Representations}
\label{subsec:inr}
To make the network scalable to large graphs, we use an embedding lookup table $\mathbf{X}_e\in \mathbb{R}^{d_e \times |M+N|}$ to map each node to a low-dimensional vector $\mathbf{x}_e \in \mathbb{R}^{d_e}$, where $d_e \ll |M+N|$. $\mathbf{X}_e$ is trained end-to-end along with the network. 
However, naively replacing the one-hot vectors with embeddings, fails to tackle the cold start problem because we cannot set the embeddings of nodes that are not seen in the training phase. 

So, to generalize the embedding learning technique to new nodes and preserve the high prediction accuracy, we take an approach of masking some percentage of the input nodes at random and then reconstructing the clean node embeddings.
Like Devlin et al.~\shortcite{devlin2018bert}, in each training batch, we mask $P_m$ percentage, say 20\%, of the whole input nodes at random. Then we reconstruct these masked embeddings. For the masked nodes, they perform the following choices: (1) with probability $p_z$, we set the node embeddings to be zero; and (2) with the remaining probability, we keep the node unchanged. 

Training with the masked embedding mechanism has two advantages. First, it can learn embeddings for nodes that are not observed in the training phase. In a cold start scenario, we initialize the embeddings of new nodes to be zero and gradually refine the estimated embeddings by multiple blocks of GCN encoder-decoders. For instance, the first block predicts the embedding of the new node by leveraging the neighborhood data (or node attributes, if available). Then the predicted embedding is fed to the second block to predict ratings and a refined embedding. The rating and embedding prediction losses are jointly optimized. Thus, STAR-GCN can solve the cold-start issue by iteratively refining the embeddings and is fundamentally different from GC-MC.
Second, STAR-GCN leads to improvement in the transductive setting. In the training phase, part of the node embeddings are masked and the network is asked to reconstruct these masked embeddings, which requires the network to encode the relationships between users and items effectively. Thus, the reconstruction loss acts as a multi-task regularizer that improves the performance of the primary rating prediction task.

When external node features are available, they are first processed via a separate network and then concatenated with the node embeddings. The feature vector $\mathbf{f}$ is mapped to a fixed size vector $\mathbf{x}_f \in \mathbb{R}^{d_f}$ using a two-layer feedforward neural network, i.e., $\mathbf{x}_f = \mathbf{W}_{2} \alpha(\mathbf{W}_{1} \mathbf{f})$, where both layers have an output dimension $d_f$.
Now the input node vector $\mathbf{x}$ becomes $[\mathbf{x}_e; \mathbf{x}_f]$ and the input dimension $d_{in} = d_e + d_f$ rather than $\mathbf{x}_e$ when the content information is not available.

\subsection{Graph Encoder and Decoder}
\label{subsec:gae}
The graph encoder transforms an input node vector $\mathbf{x}\in \mathbb{R}^{d_{in}}$ into hidden state $\mathbf{h}$ of size $d_h$ by aggregating neighboring information of different rating levels, i.e., $\mathbf{h}=\text{Enc}(\mathbf{x})$.
We choose the encoder to be the multi-link GCN aggregator in GC-MC, which is formulated in Eq.(\ref{eq:gcmc}).
A decoder maps the structural-encoded node representation $\mathbf{h}$ to a $d_{in}$-dimensional reconstructed embedding vector $\mathbf{\hat{x}}$, i.e., $\mathbf{\hat{x}} = \text{Dec}(\mathbf{h})$. We use a two-layer feedforward neural network as a decoder, i.e., $\mathbf{\hat{x}} = \mathbf{W}_{4} \alpha(\mathbf{W}_{3} \mathbf{h})$, where the output dimensions are both $d_{in}$.

STAR-GCN is a general framework with a stack of GCN encoder-decoder blocks. Any variant of GCNs, e.g., GraphSAGE~\cite{hamilton2017inductive} and GAT~\cite{velivckovic2018graph}, can be used as an encoder or decoder in STAR-GCN.
Our graph encoder-decoder is different from the graph auto-encoder model~\cite{kipf2016variational} mainly in the role of the decoder. Our decoder is to recover the initial input node vectors, while their decoder is a task-specific classifier to produce predictions. Another difference is that our graph aggregator considers different link types~\cite{schlichtkrull2018modeling}, whereas their aggregator only models single link type.

\begin{figure}
    \centering
    \begin{subfigure}[b]{0.47\linewidth}
        \centering
        \includegraphics[width=0.75\textwidth]{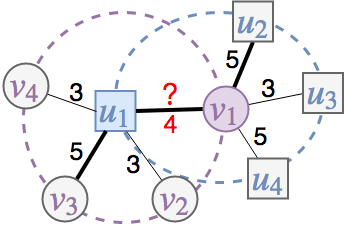}
        \caption{The original node and edge input to an aggregator, including the ground truth label 4.}
        \label{fig:data_leakage1}
    \end{subfigure}
    ~
    \begin{subfigure}[b]{0.48\linewidth}
        \centering
        \includegraphics[width=0.74\textwidth]{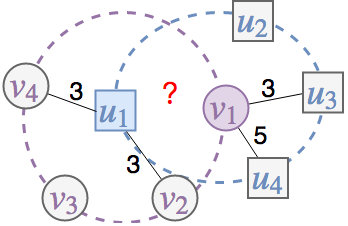}
        \caption{The nodes and edges to be aggregated after removing the sampled edges.}
        \label{fig:data_leakage2}
    \end{subfigure}
    \vspace{-0.2cm}
    \caption{The graph aggregation operator for $u_1$ and $v_1$ when predicting the rating value between $u_1$ and $v_1$. The dash circles include the nodes and edges to be aggregated. The links in bold lines are the sampled edges (rating pairs) to be trained. (a) illustrates the leakage issue. (b) gives a solution by removing the sampled edges.}
    \label{fig:dataleakage}
    \vspace{-0.3cm}
\end{figure}

\subsection{Loss}
\label{subsec:lt}
Suppose there are $L$ blocks, the loss function is expressed as
\vspace{-0.1cm}
\begin{equation}
    \begin{aligned}
        \mathcal{L} &= \sum_{\ell=1}^L (\mathcal{L}_t^{(\ell)} + \lambda^{(\ell)} \mathcal{L}_r^{(\ell)}),
    \end{aligned}
\vspace{-0.1cm}
\end{equation}
where $\mathcal{L}_t^{(\ell)}$ is a supervised task-specific loss, i,e., the rating prediction loss, and $\mathcal{L}_r^{(\ell)}$ is a reconstruction loss from the $\ell$-th block. $\lambda^{(\ell)}$ is a constant weighting factor. In the following description, we omit the layer superscript for brevity.

Suppose we have a batch of sampled edges $\mathcal{E}_b$ and two sets of masked nodes $\mathcal{U}_m$ and $\mathcal{V}_m$, the specific losses are
\begin{align*}
    \mathcal{L}_t &= \frac{1}{|\mathcal{E}_b|} \sum_{(u_i,v_j)\in \mathcal{E}_b}(r_{i,j} - \mathbf{u}_i ^T \mathbf{v}_j)^2, \\
    \mathcal{L}_r &= \frac{1}{2|\mathcal{U}_m|} \sum_{u\in \mathcal{U}_m}||\mathbf{x}_u - \hat{\mathbf{x}}_u||^2+\frac{1}{2|\mathcal{V}_m|} \sum_{v\in \mathcal{V}_m}||\mathbf{x}_v - \hat{\mathbf{x}}_v||^2.
\end{align*}
For the rating prediction loss, $\mathbf{u}_{i}$ and $\mathbf{v}_{j}$ are generated by linear transforms with the output of a graph encoder, i.e, $\mathbf{u}_{i} = \mathbf{W}_r^u \mathbf{h}_{u_i}$ and $\mathbf{v}_{j} = \mathbf{W}_r^v \mathbf{h}_{v_j}$ with user or item-specific matrix parameters $\mathbf{W}_r^u$, $\mathbf{W}_r^v \in \mathbb{R}^{d_r \times d_o}$. We train the overall models end-to-end with backpropagation.

During the inference period, a $L$-block STAR-GCN can produce $L$ predictions. Generally, we take the prediction from the last block as a final result.

\subsection{Training by Avoiding a Leakage Issue}
\label{subsec:issue}
When training the GCN-based models for rating prediction, we discover a training label leakage issue. The ground-truth training labels are involved in the model input, which significantly degrades the testing prediction performance. Specifically, suppose we have a training input $x$ and a label $y$, the model should be trained as $y=f_{\theta}(x)$. However, if the training label leakage occurs, the model becomes $y=f_{\theta}(x,y)$ resulting in an overfitting problem. The label leakage problem is a GCN specific issue because of the neighborhood aggregation operator. The user-item rating values in the training set are used to construct edges of the bipartite graph. Thus, when we utilize the neighboring data to update a node's representation, the rating values, which we need to predict, are included in the graph structure. This causes the label leakage issue.  As in Figure~\ref{fig:data_leakage1}, $y$ is the edge value $4$ between $u_1$ and $v_1$, which is taken as the input to a graph aggregator.

To avoid the leakage issue, we provide a \emph{sample-and-remove} training strategy. At each iteration, we sample a fixed size batch of rating pairs and remove the sampled pairs (edges) from the training graph before we start training the model. As in Figure~\ref{fig:data_leakage2}, the sampled edges with bold links are removed from the graph when we aggregate neighbors. After avoiding this leakage issue, the network shows a substantial boost in performance. The compared results of the leakage issue are given in Table 3.

\begin{table}[t]
    \centering
    \caption{Statistics of the datasets. `$D_{U}$' and `$D_{V}$' are the input feature dimension of users and items, respectively.}
    \vspace{-0.2cm}
    \footnotesize
    \begin{tabular}{m{1.3cm} |m{0.4cm}|m{0.5cm}|m{0.7cm}|m{0.7cm}|m{1.0cm}|m{0.8cm}}
        \hline
                &$D_{U}$&$D_{V}$& \#U& \#V & $\mathcal{R}$ & \#R\\ \hline
        Flixster& 3K & 3K & 2,341 & 2,956 & 0.5,1,..,5& 26,173    \\
        Douban  & 3K & -     & 2,999 & 3,000 & 1,..,5    & 136,891    \\ 
        ML-100K & 23    & 320   & 943   & 1,682 & 1,..,5    & 100K     \\
        ML-1M   & 23    & 320   & 6,040 & 3,706 & 1,..,5    & 1M \\
        ML-10M  & -     & 321   & 69,878& 10,677& 0.5,1,..,5& 10M\\ 
        \hline
    \end{tabular}
    \label{tb:dataset}
    \vspace{-2mm}
\end{table}

\begin{table*}[t]
    \centering
    \caption{Comparison of test RMSE scores for transductive rating prediction.}
    \small
    \begin{tabular}{l|l|l|l|l|l}
        \hline
        & Flixster & Douban & ML-100K & ML-1M & ML-10M \\ \hline  \hline
        BiasMF~\cite{koren2009matrix}           &-&-& 0.917  & 0.845 & 0.803 \\ 
        NNMF~\cite{dziugaite2015neural}         &-&-& 0.907  & 0.843 &  -         \\
        I-AUTOREC~\cite{sedhain2015autorec}     &-&-& -      & 0.831 & 0.782     \\
        GRALS~\cite{rao2015collaborative}       & 1.245 & 0.833 & 0.945 & - & - \\ 
        CF-NADE~\cite{Zheng2016neural}          &-&-& -      & \textbf{0.829} & 0.771    \\
        Factorized EAE~\cite{hartford2018deep}  &-&-& 0.910 & 0.860 & - \\
        sRMGCNN~\cite{monti2017geometric}       & 0.926 & 0.801 & 0.929 & - & - \\
        GC-MC~\cite{berg2017graph}              & 0.917 & 0.734 & 0.910 & 0.832 & 0.777 \\
        STAR-GCN                                & \textbf{0.879}$\pm$0.0030 & \textbf{0.727}$\pm$0.0006 & \textbf{0.895}$\pm$0.0009 & 0.832$\pm$0.0016 & \textbf{0.770}$\pm$0.0001\\ \hline
    \end{tabular}
    \label{tb:exp_trans}
    \vspace{-2mm}
\end{table*}
\section{Experiments}
We conduct extensive experiments on five popular recommendation benchmarks for the transductive and inductive rating prediction tasks.
The datasets are summarized in Table~\ref{tb:dataset}.
\textbf{Flixster} and \textbf{Douban} are preprocessed and provided by Monti et al.~\shortcite{monti2017geometric}. The user and item features are the adjacency vectors of the respective user-user and item-item graphs. 
The MovieLens\footnote{Movielens~\cite{harper2016movielens}: \url{https://grouplens.org/datasets/movielens/}} datasets contain different scales of rating pairs, i.e., 100 thousand, 1 million, and 10 million. We denote them as \textbf{ML-100K}, \textbf{ML-1M}, and \textbf{ML-10M}.
For user features, we take the \emph{age} as a scalar and the \emph{gender} as a binary numerical value, and the \emph{occupation} as a one-hot encoding vector. For movie features, we concatenate the title name, release year, and one-hot encoded genres. We process title names by averaging the off-the-shelf 300-dimensional GloVe CommonCrawl word vector~\cite{pennington2014glove} of each word.

We take the commonly adopted \emph{Root Mean Squared Error} (RMSE) metric to evaluate the prediction accuracy between the ground truth $r_i$ and the predicted rating $\hat{r}_i$,
\vspace{-0.2cm}
\begin{equation}
\begin{aligned}
\text{RMSE} =  \sqrt{\frac{\sum_{i=1}^N(\hat{r}_i - r_i)^2}{N}}.
\end{aligned}
\vspace{-0.2cm}
\end{equation}

\begin{table}[tb]
    \centering
    \caption{Ablation analysis of different STAR-GCN models in transductive rating prediction. `b' denotes the \emph{block} number, `l' denotes the number of graph convolution \emph{layer}s in each block, and `-' denotes minus. `- rec.' means the model does not have the reconstruction module and `- rm.' denotes the model is trained without removing sampled edges. `recurrent' denotes that the blocks are combined by recurrence. Otherwise, the blocks are combined by stacking.}
    \footnotesize
    \begin{tabular}{m{0.5cm}|m{1.7cm}|m{0.7cm}|m{0.7cm}|m{0.7cm}|m{0.7cm}|m{0.7cm}}
        \hline
        &Models& Flix. & Dou. & ML-100K & ML- 1M & ML-10M \\ \hline \hline
        &1b2l {\tiny(- rec., - rm.)}   & 0.920 & 0.731 & 0.921 & 0.841 & 0.782 \\  
        \scriptsize{Emb.} &1b2l {\tiny(- rec.)}         & 0.893 & 0.728 & 0.899 & 0.835 & 0.778 \\  
        \scriptsize{only} &1b2l                        & 0.891 & 0.728 & 0.901 & 0.834 & 0.771 \\ 
         &2b1l {\tiny(recurrent)}     & 0.883 & \textbf{0.727} & \textbf{0.895} & 0.833 & 0.773 \\  
        &2b1l                        & \textbf{0.879} & \textbf{0.727} & 0.898 & \textbf{0.832} & 0.771 \\ \hline
        &1b2l {\tiny(- rec.,- rm.)}    & 0.917 & 0.731 & 0.920 & 0.840 & 0.782 \\  
        \scriptsize{With} &1b2l {\tiny( - rec.)}         & 0.889 & 0.727 & 0.899 & 0.835 &  0.778\\ 
        \scriptsize{Fea.} &1b2l                        & 0.887 & 0.728 & 0.901 & 0.834 & \textbf{0.770} \\ 
         &2b1l {\tiny(recurrent)}     & \textbf{0.879} & 0.728 & 0.896 & 0.833 & 0.772 \\  
        &2b1l                        & 0.880 & \textbf{0.727} & 0.896 & \textbf{0.832}& 0.771 \\ \hline
    \end{tabular}
    \label{tb:ablation}
    \vspace{-2mm}
\end{table}

\begin{table*}[t]
    \centering
    \caption{Comparison of test RMSE scores for inductive rating prediction. `- rec.' denotes the model does not have the reconstruction module and `+ fea.' means the model uses external node features. We train all models three times and report the mean scores and the standard deviation.}
    \footnotesize
    \begin{tabular}{c | l | c c c | c c c}
        \hline
        \multirow{2}{*}{Datasets}& \multirow{2}{*}{Models}  & \multicolumn{3}{c|}{Items 20\%}   & \multicolumn{3}{c}{Users 20\%}  \\  \cline{3-8}
        &  & 50\% & 30\% & 10\%     & 50\% & 30\% & 10\%  \\ \hline \hline
        \multirow{6}{*}{Douban}
        & DropoutNet                            & - & - & - & 0.797$\pm$0.002 & 0.797$\pm$0.003 & 0.797$\pm$0.001    \\
        & CDL                                   & - & - & - & 0.781$\pm$0.006 & 0.781$\pm$0.001 & 0.781$\pm$0.001 \\
        & STAR-GCN{\scriptsize(- rec.)}          & 0.734$\pm$0.001 & 0.746$\pm$0.001 & 0.777$\pm$0.002 & 0.731$\pm$0.000 & 0.738$\pm$0.000 & 0.753$\pm$0.001 \\
        & STAR-GCN{\scriptsize(- rec., + fea.)}   & - & - & - & 0.731$\pm$0.002 & 0.737$\pm$0.000    & 0.753$\pm$0.001 \\
        & STAR-GCN                              & \textbf{0.725}$\pm$0.001 & \textbf{0.734}$\pm$0.001 & \textbf{0.764}$\pm$0.000 & \textbf{0.725}$\pm$0.001 & \textbf{0.731}$\pm$0.001 & 0.747$\pm$0.001 \\
        & STAR-GCN{\scriptsize(+ fea.)}          & - & - & - & \textbf{0.725}$\pm$0.002 & \textbf{0.731}$\pm$0.000 & \textbf{0.746}$\pm$0.000  \\
        \hline
        \multirow{6}{*}{ML-100K}
        & DropoutNet                             & 1.223$\pm$0.065 & 1.167$\pm$0.031 & 1.144$\pm$0.024 & 1.015$\pm$0.002 & 1.022$\pm$0.006 & 1.023$\pm$0.003 \\
        & CDL                                    & 1.083$\pm$0.009 & 1.082$\pm$0.007 & 1.082$\pm$0.007 & 1.011$\pm$0.005 & 1.013$\pm$0.006 & 1.015$\pm$0.004 \\
        & STAR-GCN{\scriptsize(- rec.)}           & 0.932$\pm$0.001 & 0.943$\pm$0.001 & 0.976$\pm$0.003 & 0.919$\pm$0.002 & 0.933$\pm$0.001 & 0.949$\pm$0.001\\
        & STAR-GCN{\scriptsize(- rec., + fea.)}    & 0.928$\pm$0.002 & 0.941$\pm$0.002 & 0.977$\pm$0.004 & 0.916$\pm$0.005 & 0.931$\pm$0.004 & 0.951$\pm$0.005\\
        & STAR-GCN                               & 0.919$\pm$0.001 & \textbf{0.926}$\pm$0.000 & \textbf{0.954}$\pm$0.001 & \textbf{0.907}$\pm$0.004 & \textbf{0.917}$\pm$0.005 & 0.937$\pm$0.005 \\
        & STAR-GCN{\scriptsize(+ fea.)}           & \textbf{0.918}$\pm$0.002 & \textbf{0.926}$\pm$0.002 & 0.956$\pm$0.000 & \textbf{0.907}$\pm$0.002 & \textbf{0.917}$\pm$0.001 & \textbf{0.936}$\pm$0.004\\
        \hline
        \multirow{6}{*}{ML-1M}
        & DropoutNet                            & 1.169$\pm$0.120 & 1.134$\pm$0.034 & 1.256$\pm$0.128 & 1.002$\pm$0.001 & 1.005$\pm$0.005 & 1.003$\pm$0.001 \\
        & CDL                                   & 1.068$\pm$0.009 & 1.069$\pm$0.009 & 1.068$\pm$0.009 & 0.974$\pm$0.000 & 0.975$\pm$0.000 & 0.974$\pm$0.000 \\
        & STAR-GCN{\scriptsize(- rec.)}          & 0.862$\pm$0.001 & 0.872$\pm$0.004 & 0.903$\pm$0.004 & 0.859$\pm$0.002 & 0.868$\pm$0.001 & 0.891$\pm$0.001 \\
        & STAR-GCN{\scriptsize(- rec., + fea.)}   & 0.861$\pm$0.002 & 0.867$\pm$0.002 & 0.910$\pm$0.006 & 0.859$\pm$0.001 & 0.869$\pm$0.001 & 0.893$\pm$0.001 \\
        & STAR-GCN                              & \textbf{0.844}$\pm$0.000 & \textbf{0.850}$\pm$0.000 & \textbf{0.876}$\pm$0.004 & \textbf{0.848}$\pm$0.001 & \textbf{0.858}$\pm$0.001 & \textbf{0.882}$\pm$0.000 \\
        & STAR-GCN{\scriptsize(+ fea.)}          & \textbf{0.844}$\pm$0.001 & 0.851$\pm$0.001 & \textbf{0.876}$\pm$0.002 & 0.849$\pm$0.001 & \textbf{0.858}$\pm$0.000 & 0.883$\pm$0.001\\
        \hline
    \end{tabular}
    \label{tb:exp_ind}
    \vspace{-2mm}
\end{table*}

\subsection{Model Architecture and Implementation Details}
\label{subsec:model_arch}
We test the overall network design with different sets of hyperparameters. The validation set determines our final design.
We regard Flixtser, Douban, and ML-100K as small datasets and ML-1M and ML-10M as large datasets. After tuning the hyperparameters, we roughly apply two sets of hyperparameters, one for small datasets and the other for large datasets.
In all models, we choose the non-linear function $\alpha(\cdot)$ as a LeakyReLU activation with the negative slope equals to 0.1. 
For the input vectors, we set the dimension of node embeddings $d_e$ to be 32 for small datasets and 64 for large datasets. When incorporating features, we take the projection dimension $d_f$ to be 8 for small datasets and 32 for large datasets. 
Regarding the masking mechanism, for transductive prediction, we randomly select $P_m$=$10$\% of the nodes for reconstruction and set $p_z$=$0$. For inductive prediction, we uniformly choose $P_m$=$40$\% nodes and mask them to be zero, i.e., $p_z$=$1$, to approximate the testing data distribution.
For encoders, the hidden size $d_a$ sets to be 250. The dimension of the output layer $d_h$ sets to be 75. We apply a dropout layer to the input of a GCN layer with a dropout rate of 0.5 for small datasets and 0.3 for large datasets.
For decoders, all the hidden sizes are fixed as the node input dimension, i.e., $d_{in}$.  
When predicting ratings, we set the projection size $d_r$ to be 64.

We train the STAR-GCN models with Adam~\cite{kingma2014adam} optimizer and use the validation set to perform learning rate decay scheduler. The initial learning rate is set to be 0.002 and gradually decreases to be 0.0005 with the decay rate of 0.5 each time the validation RMSE score does not fall in a window of 100 iterations, and the early stopping occurs for 150 iterations. The gradient normalization value clips to be no larger than 1.0. 
The training batch size is fixed to be 10K for small datasets, 100K for ML-1M, and 500K for ML-10M. We train the model three times with different random seeds, except for ML-10M training two times, and report the average test RMSE scores along with the standard deviation.

\subsection{Transductive Rating Prediction} 
All the datasets are used for transductive rating prediction evaluations, where the users and items are all observed in the training dataset. For fair comparison, we strictly follow the experimental setup of Berg et al.~\shortcite{berg2017graph}. For Douban and Flixster, we use the split test set provided by Monti et al.~\shortcite{monti2017geometric} with 10\% of rating pairs as the testing set. For ML-100K, we use the first of the five provided data split with 20\% for testing. For ML-1M and ML-10M, we randomly split the edges with 10\% for testing. In the transductive setting, we perform a thorough comparison of STAR-GCN with multiple baselines and state-of-the-art models listed in Table~\ref{tb:exp_trans}. Baseline Results are taken from Berg et al.~\shortcite{berg2017graph}.
We also conduct comprehensive ablation analysis in Table~\ref{tb:ablation}.

Table~\ref{tb:exp_trans} summarizes all results of the baselines and our STAR-GCN. The baseline scores are directly taken from Monti et al.~\shortcite{monti2017geometric} and Berg et al.~\shortcite{berg2017graph}. 
The results of STAR-GCN are produced by different variant models with a total of two graph convolutional layers, labeled as either `1b2l' or `2b1l' in Table~\ref{tb:ablation}. `1b2l' denotes a model with one block of encoder-decoder and each encoder includes a two-layer GCN, as in Figure~\ref{fig:layer_recon}. In contrast, `2b1l' indicates a model with two encoder-decoder blocks and each encoder contains a one-layer GCN, as in Figure~\ref{fig:block_recon}. In particular, `1b2l (- rec.)' indicates the model only has an encoder without the reconstruction module, as in Figure~\ref{fig:block_norecon}.
The reported RMSE scores for STAR-GCN in Table~\ref{tb:exp_trans} is the best results of different STAR-GCN models listed in Table~\ref{tb:ablation}. We note that the proposed STAR-GCN architecture achieves the state-of-the-art results on four out of five datasets. We have the following findings from Table~\ref{tb:ablation}.

\textbf{Effect of removing sampled training edges within mini-batches.}
Comparing the results of `1b2l~(- rec. - rm.)' and `1b2l~(- rec.)', we see a significant decrease in the testing RMSE scores after removing the sampled user-item pairs from the bipartite graph in each training batch. This proves the effectiveness of our sample-remove training strategy to avoid the training data leakage issue.

\textbf{Effect of reconstructing masked nodes.}
Comparison of the results of the models labeled `- rec.' with the models having no such label indicates that the models possessing the reconstruction module consistently beat the models without reconstruction utilizing the same total number of graph encoders, which proves that our reconstruction mechanism is beneficial to the final prediction performance.

\textbf{Effect of the recurrent structure.}
Comparing the results of `2b1l~(recurrent)' and `2b1l', we see that the recurrent structure can achieve competitive results with fewer parameters.

\textbf{Effect of incorporating features.}
By comparing the results from the same models with and without features, we note that combining external node features does not always produce better performance.

\subsection{Inductive Rating Prediction}
We conduct inductive rating prediction experiments on three datasets, Douban, ML-100K, and ML-1M. For each dataset, we keep 20\% of user (or item) nodes as the testing nodes and remove them from the training graph. Then, we choose a fraction of ratings linked with the testing nodes as the edges that are observed in the testing phase. Ratings that are not chosen by this step are kept as the testing data. The predictor never sees these 20\% nodes and needs to rely on the observed links (together with the node features, if available) to predict ratings. We conduct experiments on three different fractions, 50\%, 30\%, and 10\%, which means that there are 50\%, 30\%, and 10\% ratings linked with new nodes available to the predictor in the testing phase. Intuitively, the more edges we access, the more information we have, and the better the performance will be. We implement two baseline models, CDL~\cite{wang2015collaborative} and DropoutNet~\cite{volkovs2017dropoutnet}, and compare some variants of our STAR-GCN architecture. We use the `2b1l` versions of STAR-GCN in this task.

The testing RMSE scores are listed in Table~\ref{tb:exp_ind}. We find a worse performance tendency when the predictor accesses fewer neighboring edges for the new users/items in the testing phase. 
We show that our STAR-GCN model produces significantly better results than two baselines. 
Moreover, comparing the results of the models with and without reconstruction modules, we find that the reconstruction mechanism plays a crucial role in improving the final performance.
An interesting observation is that incorporating content information for new nodes is not always beneficial to the final results. The reason may be that our reconstructed node embeddings have already contained enough information for accurate predictions, which proves that our STAR-GCN can effectively solve the cold start problem using structural information.

\section{Conclusion and Future Work}
We introduce a new GCN-based architecture and apply it to transductive and inductive rating prediction. Our STAR-GCN achieves the state-of-the-art results in both tasks. Our architecture is generic and can be used in other applications, such as abnormal behavior detection, spatiotemporal forecasting~\cite{shi2018machine}, thread popularity prediction~\cite{ChanK18}, and so on~\cite{gao2018difficulty}. Moreover, we discover a training label leakage issue when implementing GCN-based models for rating prediction tasks. The discovery should serve as a reminder of later research.
In the future, we plan to improve our STAR-GCN to handle heterogeneous graphs with diverse node types for better simulating real-life scenarios and to integrate ranking algorithms~\cite{SuKL17} to solve other recommendation tasks.

\section{Acknowledgement}
The work described in this paper was partially supported by the Research Grants Council of the Hong Kong Special Administrative Region, China~(No.~CUHK~14208815 of the General Research Fund) and Meitu~(No.~7010445).

\newpage
\bibliographystyle{named}
\bibliography{graph19}

\end{document}